\documentstyle[prl,multicol,aps,epsfig,prb]{revtex}
\begin{document}
\draft 

% Title and header ------------------------------------------------------------
\title{Nonequilibrium dynamical mean-field theory of a strongly correlated system}
\author{P.~Schmidt and H.~Monien} 
\address{Physikalisches Institut, Universit\"at Bonn, 
         Nu{\ss}allee 12, D-53115 Bonn, Germany}
 
%\twocolumn[
\date{\today} \maketitle\widetext 
% Abstract --------------------------------------------------------------------
\begin{abstract}
\begin{center}
  \parbox{14cm}{ 
We present a generalized dynamical mean-field approach for the  nonequilibrium physics of a
strongly correlated system in the presence of a time-dependent external field. The Keldysh Green's function formalism is used to 
study the nonequilibrium problem. We derive a closed set of self-consistency equations in the case of a 
driving field with frequency $\Omega$ and wave vector ${\bf q}$. We present numerical results for the 
local frequency-dependent Green's function and the self-energy for different values of the field amplitude in the case of a 
uniform external field using the iterated perturbation theory. In addition, an expression for the frequency-dependent 
optical conductivity of the Hubbard model with a driving external field is derived.}
\end{center}
\end{abstract}
\pacs{\hspace{1.9cm} 
      }%] %75.10.Jm, 75.40.Mg, 75.50.Ee, 71.38.+i, 63.20.Kr } ]
%\narrowtext
%
% Introduction ----------------------------------------------------------------
%\section{Introduction}
%
\begin{multicols}{2}

The study of nonequilibrium many-body phenomena is a field of great and still growing interest in today's condensed matter physics.  
Recent pump-probe experiments on transition metal oxides show unusual effects, e.g. a dramatic 
change in the optical transmission \cite{OA}, which require a theoretical description of driven quantum systems.
A basic model for theoretical investigations of strongly correlated fermion systems is the Hubbard model.
Within the scope of the dynamical mean-field theory (DMFT) \cite{GKII}, a very detailed analysis of the spectral properties of the 
Hubbard model became possible. The DMFT approach takes into account local correlations as well as high energy excitations
and provides a useful tool for calculating linear response functions of strongly correlated fermion systems \cite{PCJI,PCJII}.
%The basic approximation of the DMFT is the assumption of a local self-energy, which can be justified in the limit of huge 
%lattice connectivity \cite{}. Within DMFT the original lattice problem is mapped onto an effective impurity problem, which can be 
%solved numerically, using e.g. quantum-monte-carlSo simulations, iterated perturbation theory \cite{}.

In this paper, we present a generalization of the DMFT using the Keldysh formalism to study a Hubbard model under the 
influence of a periodic external perturbation.
\begin{eqnarray}
 \label{eq:hamiltonian}
 H & = &-\sum\limits_{\left<ij\right>\sigma}^{} t_{ij}\left(c_{i\sigma}^{\dag} c_{j\sigma}^{}
 + c_{j\sigma}^{\dag} c_{i\sigma}^{}\right) + U\sum\limits_{i}^{} n_{i\uparrow}n_{i\downarrow} \nonumber\\
 & &  + \sum_{i\sigma} V_i(t)c_{i\sigma}^{\dag}c_{i\sigma}^{},
\end{eqnarray}
where $c_{i\sigma}^{\dag} (c_{i\sigma}^{})$ creates (annihilates) a fermion with spin 
$\sigma$ at site $i$ and $n_{i\sigma} = c_{i\sigma}^{\dag}c_{i\sigma}^{}$ is the local density operator.
The bare half-filled single-band Hubbard model describes nearest-neighbor hoppings of electrons on a lattice, 
interacting with each other through the on-site Coulomb interaction.
The additional explicitly time-dependent term in the Hamiltonian describes the external driving field with frequency $\Omega$, wave vector 
${\bf q} = (q,...,q)$ and arbitrary field amplitude $A$.
\begin{equation}
V_i(t) = A\cos\left(\Omega t-{\bf qR}_i\right)
\end{equation}
In particular we are interested in the small-$q$ limit. This Hamiltonian describes photodoping of a strongly correlated material,
since the photon wavelength is large compared to the lattice spacing. For a uniform external field ($q=0$) the time-dependent perturbation term commutes 
with the Hubbard Hamiltonian, and this problem can be solved quite easily, as we will see later.
%
%For large commensurate $q$, e.g. $q=\pi$, the d-dimensional lattice can be divided into $2\pi/q$ sublattices which 
%have different local properties.
%Performing the limes of a small wave vector is numerically limited by the increasing number of sublattices.

For large commensurate wave vectors, e.g. $q=\pi$, the $d$-dimensional lattice can be divided into cluster of $N=2\pi/q$ sites. 
%with periodic boundary condition. 
The local properties depend on the chosen site ${\bf R}$ of this cluster. 
%In this case the method described below is based on the dynamical cluster approximation (DCA) \cite{HTJ}, a generalization of the DMFT. 
%The original driven Hubbard model is mapped onto a self-consistent embedded cluster instead of single impurity.
%Decreasing the wave vector require solving cluster problems of increasing size, which is numerically limited because of an exponentially
%growing Hilbert space.
Performing the limit of a small wave vector is numerically limited by the increasing number of sites in the cluster.
%-------------------------------------------------------------------------------
A theory which sys\-te\-ma\-ti\-cally includes nonlocal correlations to the DMFT is the dynamical cluster approximation (DCA)
\cite{HTJ}, where the original model is mapped onto a self-consistent embedded cluster instead of a single impurity.
%------------------------------------------------------------------------------

The main technique for the treatment of non\-equi\-li\-bri\-um effects is the Keldysh Green's function formalism \cite{K,RS}.
%based on a contour-ordered Green's function which provides information about the quantum states of the system as well as about 
%their occupation out of equilibrium within a density matrix formulation.
%Nonequilibrium expectation values are defined with respect to the states of the system in the remote past before the driving
%field has been switched on \cite{RS}. The entire time-dependence of the field is treated nonperturbatively.
%Man koennte die beiden obigen Zeilen weglassen.
%
The usual time-ordering is replaced by a contour-ordering along the Keldysh contour, which follows the real-time axis from 
$-\infty$ to $+\infty$ and back to $-\infty$. 
The resulting Green's functions matrix $G^{\alpha\beta}$, where $\alpha, \beta\in\{\pm 1\}$ specifies which time argument 
lies on the upper $(+)$, respectively lower $(-)$ branch of the Keldysh contour, defines three linearly independent components 
which provide information about the quantum states of the system (retarded/advanced Green's function) as well as about their occupation 
out of equilibrium within a density matrix formulation (Keldysh component).

The external field is periodic in time and in each spatial coordinate. 
%In the commensurable case the d-dimensional lattice can be divided into cells of size $(2\pi/q)^{d}$ with periodic boundary condition. 
%DAS IST DIE EINHEITSZELLE, SCHUSSELCHEN!
Thus, according to a generalized Floquet theorem \cite {PH}, energy and momentum are conserved up to multiples of the external frequency 
and wave vector, respectively. 
We define a double Fourier transformation, expanding the Green's function in its Floquet modes
\begin{eqnarray}
G_{ij}^{\alpha\beta}(t,t')  & = &
	  \sum_{{\bf k},n}\int\frac{d\omega}{2\pi}e^{-i\omega t}e^{i\left(\omega-n\Omega\right)t'}  \nonumber\\
	& \times & e^{i{\bf kx}_{i}}e^{-i\left({\bf k}-n{\bf q}\right){\bf x}_{j}}G_{0n}^{\alpha\beta}\left({\bf k},\omega\right). 
\end{eqnarray}
%
%For describing photoexcitation/photodoping, we are interested in the case of small wave vector since the lattice spacing in 
%usually large compared to the photon wave vector. In the case of a uniform field (q=0) the perturbation commutes with the
%Hubbard Hamiltonian and the problem can be solved quite easily using DMFT. For $(q\not=0)$ the Hamiltonian contains a spatial 
%fluctuating term. 
%For of a uniform field (q=0) the perturbation commutes with the Hubbard Hamiltonian and the problem can be solved quite easily using DMFT.
%For large q (e.g. the nesting vector $(\pi,...,\pi)$ ) which corresponds to short-range fluctuations the system can be solved
%in terms of the dynamical cluster approximation. One has to average over the 
%
In the noninteracting case ($U=0$) and for small  $q$, it is possible to calculate the Green's function analytically by solving the equation 
of motion. The weighting coefficient of the $n$-th energy sideband turns out to be the Bessel function 
$J_n\left(A/\Omega_q({\bf k})\right)$ with $\Omega_q({\bf k})=\Omega-{\bf qv_{k}}$.
Thus, the noninteracting retarded/advanced Green's function is given by  
\begin{eqnarray}
% g^{\mbox{\tiny ret/av}}_{{\bf k},{\bf k}-n{\bf q}}\left(\omega, \omega-n\Omega\right) 
 g^{\mbox{\tiny ret/av}}_{0n}\left({\bf k}, \omega\right) 
	& = & \sum_{m} \frac{J_m(A_{q}({\bf k})) J_{m-n}(A_{q}({\bf k}))}
	{\omega-\epsilon_{{\bf k}}-m\Omega_{q}({\bf k})\pm i0^+}
\end{eqnarray}
where $A_q({\bf k}) = A/\Omega_q({\bf k})$.
The noninteracting Keldysh component reads
\begin{eqnarray}
\label{eq:g_keldI}
% g_{{\bf k},{\bf k}-n{\bf q}}^{\mbox{\tiny keld}}(\omega,\omega-n\Omega)
 g^{\mbox{\tiny keld}}_{0n}\left({\bf k}, \omega\right) 
 	& = & -2\pi i\sum_{m} \left(1-2f\left(\epsilon_{{\bf k}-m{\bf q}}\right)\right)\nonumber\\ 
	& & \times J_m(A_{q}({\bf k}))J_{m-n}(A_{q}({\bf k}))\nonumber\\
	& & \nonumber\\
	& & \times\delta\left(\omega-\epsilon_{{\bf k}}-m\Omega_{q}({\bf k})\right). 
\end{eqnarray}
The diagrammatic expansion of the Keldysh Green's function matrix in the Coulomb parameter $U$ is mainly identical to Feynman theory \cite{RS}. 
The only difference arises from the summation over the two time branches. 
%The resulting Dyson equation reads
%
%\begin{eqnarray}
%\label{eq:ret/avDyson}
%G^{\alpha\beta}_{0n}\left({\bf k},\omega\right) & = & g^{\alpha\beta}_{0n}\left({\bf k},\omega\right) \nonumber\\
%	& & + \sum_{\gamma\delta,lm}\gamma\delta g^{\alpha\gamma}_{0l}\left({\bf k},\omega\right) 
%	\Sigma^{\gamma\delta}_{{\bf q}lm}\left(\omega\right) G^{\delta\beta}_{mn}\left({\bf k},\omega\right) 
%\end{eqnarray}
%
%

We solve the driven Hubbard model using the dy\-na\-mi\-cal mean-field approximation. 
Within DMFT the ori\-gi\-nal lattice problem is mapped onto an effective impurity problem, which can be solved numerically, 
using e.g. quantum-Monte-Carlo simulations or iterated perturbation theory (IPT) \cite{GKI,MHII}. 
The local dynamics at a specific site ${\bf R}$ can be regarded as the interaction of the degrees of freedom 
at this site with an external bath created by all other degrees of freedom on other sites.
Due to the spatial fluctuations for nonzero external wave vector the local properties depend on the chosen site ${\bf R}$ of the cluster.
%
%Due to the spatial periodic fluctuations for nonzero $q$ the lattice can be divided into cluster of $2\pi/q$ sites and 
%the local properties depend on the chosen site ${\bf R}$ of the cluster. %in the case of the driven Hubbard model.

The key approximation of the DMFT is the assumption of a local self-energy, which can be justified in the limit of large
lattice connectivity \cite{WMI,MHI}. For the bare Hubbard model momentum conservation can be disregarded at each internal interaction 
vertex in the limit of infinite connectivity. %i.e. the self-energy of the bare Hubbard model becomes purely local \cite{}, 
An analogous argument for the driven Hubbard model yields that at internal vertices momentum conservation is automatically 
fulfilled up to multiples of ${\bf q}$.
Thus, each internal propagator $G^{\alpha\beta}_{n_1n_2}({\bf k},\omega)=G^{\alpha\beta}_{{\bf k}-n_1{\bf q},{\bf k}-n_2{\bf q}}
(\omega-n_1\Omega, \omega-n_2\Omega)$ can be replaced by $G^{\alpha\beta}_{0n}(\omega) = 
\sum_{\bf k}G^{\alpha\beta}_{q,0n}({\bf k},\omega)$ which is a ``staggered'' average over the lattice sites ${\bf R}$ in the cluster:
$\sum_{{\bf R}}e^{-in{\bf qR}}G_{{\bf R}}(\omega)$.
The resulting Dyson equation reads
\begin{eqnarray}
\label{eq:Dyson}
G^{\alpha\beta}_{q,0n}\left(\omega\right) & = & g^{\alpha\beta}_{q,0n}\left(\omega\right) \nonumber\\
 & + & \sum_{\gamma\delta,lm}\gamma\delta g^{\alpha\gamma}_{q,0l}\left(\omega\right) 
	\Sigma^{\gamma\delta}_{q,lm}\left(\omega\right) G^{\delta\beta}_{q,mn}\left(\omega\right). 
\end{eqnarray}

The mean-field description of the driven Hubbard model is determined by the effective Weiss field 
${\cal{G}}_{\bf R}(t_\alpha,t'_\beta)$, depending on the chosen site ${\bf R}$ of the cluster.
%
%by the following single-site effective action, depending on the chosen site ${\bf R}$
%
%\begin{eqnarray}
%\label{SeffII}
% S^{\mbox{\tiny eff}}_{{\bf R}}
%	& = & \sum_{\alpha\beta}\alpha\beta\int\limits_{-\infty}^{\infty}dt_{\alpha}\int\limits_{-\infty}^{\infty}dt'_{\beta}
%	\sum_\sigma\bar\psi_{{\bf R}\sigma}\left(t_\alpha\right)\nonumber\\
%	& & \nonumber\\	
%	& & \quad\times{\cal{G}}^{-1}_{\bf R}\left(t_\alpha,t'_\beta\right)\psi_{{\bf R}\sigma}\left(t'_\beta\right) \nonumber \\
%	& & \nonumber \\
%	& & - U \sum_{\alpha}\alpha\int\limits_{-\infty}^{\infty}dt_{\alpha}n_{{\bf R}\uparrow}\left(t_\alpha\right)
%	n_{{\bf R}\downarrow}\left(t_\alpha\right). 
%\end{eqnarray}
% 
The self-consistency condition is obtained by integrating out all degrees of freedom except those of the selected site ${\bf R}$.
Assuming a semicircular bare density of states (Bethe lattice) one obtains the self-consistency condition for the Fourier transformed 
Weiss field, which is a matrix equation in energy and momentum sidebands and also in Keldysh space
\begin{eqnarray}
\label{eq:SelfConsistency}
\left[{\cal G}_{q}^{-1}\right]^{\alpha\beta}_{nm}(\omega) 
	& = & \alpha\delta_{\alpha\beta}\bigg[\left(\omega-n\Omega\right)\delta_{nm}\bigg.\nonumber\\
	& & -\left.\frac{A}{2}\left(\delta_{nm+1}+\delta_{nm-1}\right)\right]\nonumber\\
	& & \nonumber\\
	& & -\alpha\beta t^2 \cos((n-m)q)G^{\alpha\beta}_{q,nm}(\omega).
\end{eqnarray}
%
%(\ref{SeffII}) and (\ref{eq:SelfConsistency}) form a closed set of self-consistency equations.
%

We used the iterated perturbation theory (IPT) \cite{GKI,MHII} to solve the mean-field equations numerically for the half-filled 
driven Hubbard model.
The basic idea of the IPT method is to perform a skeleton expansion of the self-energy up to second order in the interaction $U$: 
\begin{eqnarray}
\label{eq:Sigma0n}
 \Sigma_{q,0n}^{\gamma\delta}(\omega) & = & \gamma\delta \ U^2 \int\limits_{-\infty}^{\infty}\frac{d\omega_1}{2\pi} 
	\int\limits_{-\infty}^{\infty}\frac{d\omega_2}{2\pi} \sum_{ml} 
	{\cal G}_{q,l,0}^{\delta\gamma}\left(\omega_1\right) \nonumber\\
	& \times &{\cal G}_{q,0,n+m+l}^{\gamma\delta}\left(\omega_1+\omega_2\right) 
	{\cal G}_{q,0,-m}^{\gamma\delta}\left(\omega-\omega_2\right).
\end{eqnarray}
%
%where $G^{\alpha\beta}_{q,0n}(\omega) = \sum_{\bf k}G^{\alpha\beta}_{q,0n}({\bf k},\omega)$ denotes the local Green's function.
%
The matrix $G_{q,nm}$ is then obtained from the Dyson equation (\ref{eq:Dyson}) with the noninteracting $g_{q,nm}$
replaced by the Weiss field ${\cal G}_{q,nm}$. The self-consistency loop is closed by calculating a new Weiss field 
from equation (\ref{eq:SelfConsistency}).

In our numerical calculation, we consider the special case of a uniform external field $(q=0)$, where the time-dependent perturbation term commutes 
with the bare Hubbard Hamiltonian and therefore the weighting coefficients of the energy sidebands are Bessel functions in the interacting 
case, too.

\begin{figure}[t]  
\begin{center}
\epsfig{file=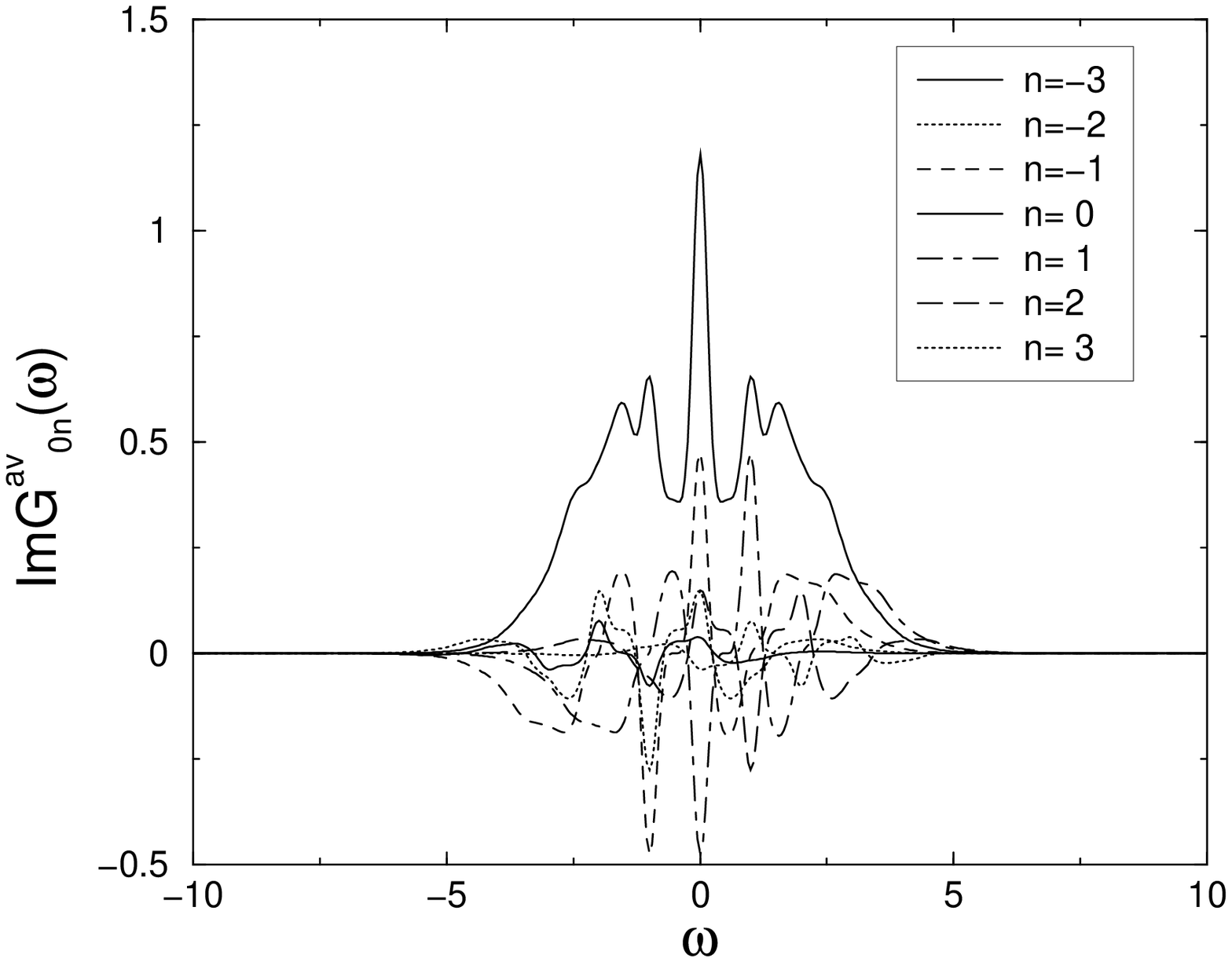, height=5cm}
\caption[]{$\mbox{Im}G_{0n}^{\mbox{\tiny av}}(\omega)$ for $U=3$, $A=1$, three sidebands.}
\label{figure:DrivenImGA1sidebands3}
\end{center}
\end{figure}

For $A=1$ and $A=2$ (we set $\Omega=1$), we restricted our considerations to the main energy 
band and the first three sidebands, i.e. we calculated the Fourier components $G_{0n}(\omega)$ with $n=-3,..,3$ 
(see Fig. \ref{figure:DrivenImGA1sidebands3}). For $A=3$, we considered five sidebands: $n=-5,..,5$.
These limitations of the number of sidebands are reasonable, since for a given $A$ the Bessel functions $J_n(A)$ 
fall off very fast with increasing $n$. This property strongly depends on the magnitude of the field: we have to take into 
account an increasing number of sidebands while enhancing the field amplitude.

As an example, the Fourier components of the spectral density for $A=1$ are given in Fig. \ref{figure:DrivenImGA1sidebands3}.  
From an inverse Fourier transformation of the advanced Green's function, we obtained that the time-dependent advanced function 
satisfies the usual causality condition.

\begin{figure}[t]
\begin{center}
\epsfig{file=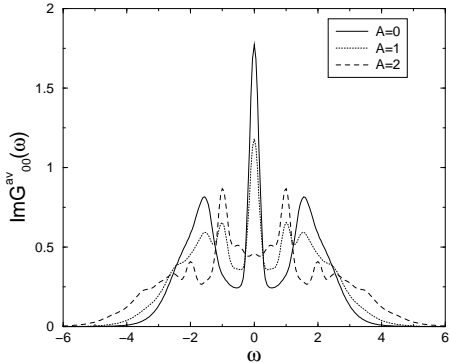, height=5cm}
\caption[]{$\mbox{Im}G_{00}^{\mbox{\tiny av}}(\omega)$ for $U=3$ and $A=0,1,2$, obtained by IPT.}
\label{figure:DrivenImGA035}
\end{center}
\end{figure}

In the presence of the uniform external field, we define an effective spectral density, given by 
\begin{equation}
 \rho^{\mbox{\tiny eff}}(\epsilon) = \frac{1}{\pi}\mbox{Im}G_{00}^{\mbox{\tiny av}}(\omega) 
	= \sum_{l}J_l^2(A)\frac{1}{\pi}\mbox{Im}\tilde{G}(\omega-l\Omega),
\end{equation}
where $\tilde{G}$ is the local Green's function for $A=0$.
Fig. \ref{figure:DrivenImGA035} shows how the effective spectral density depends on the field amplitude. 
In comparison with the usual three peak structure in the case of no external field (lower and upper Hubbard band and the 
quasiparticle peak), this function exhibits a more complex structure for $A\neq 0$ due to the sidebands, weighted by the Bessel 
functions. 
%In Fig. \ref{figure:DrivenImGA035}, we observe for $A=1$, for example, the first sideband of the quasiparticle
%peak at $\pm\hbar\Omega$ and in addition, one sideband of each Hubbard band. 
The number of occuring sidebands depends strongly on $A$: with increasing $A$, more sidebands become significant in the 
effective spectral density. 
%For $A=3$, we observe three sidebands of the quasiparticle peak.
An additional phenomenon is the strong suppression of the main quasiparticle peak for certain field amplitudes.

\begin{figure}[t]
\begin{center}
\epsfig{file=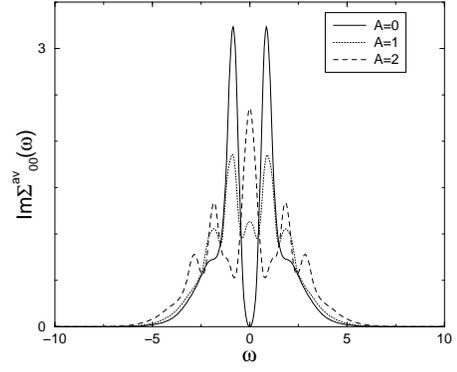, height=5cm}
\caption[]{$\mbox{Im}\Sigma_{00}^{\mbox{\tiny av}}(\omega)$ for $U=3$ and $A=0,1,2$, obtained by IPT.}
\label{figure:DrivenImSigmaA012}
\end{center}
\end{figure}

The effect of the external field on the imaginary part of the advanced self-energy on the field amplitude is shown in
Fig. \ref{figure:DrivenImSigmaA012}. For $A\neq 0$, we observe the sidebands of the typical two-peak structure 
in the equilibrium case.

\begin{figure}[t]
\begin{center}
\epsfig{file=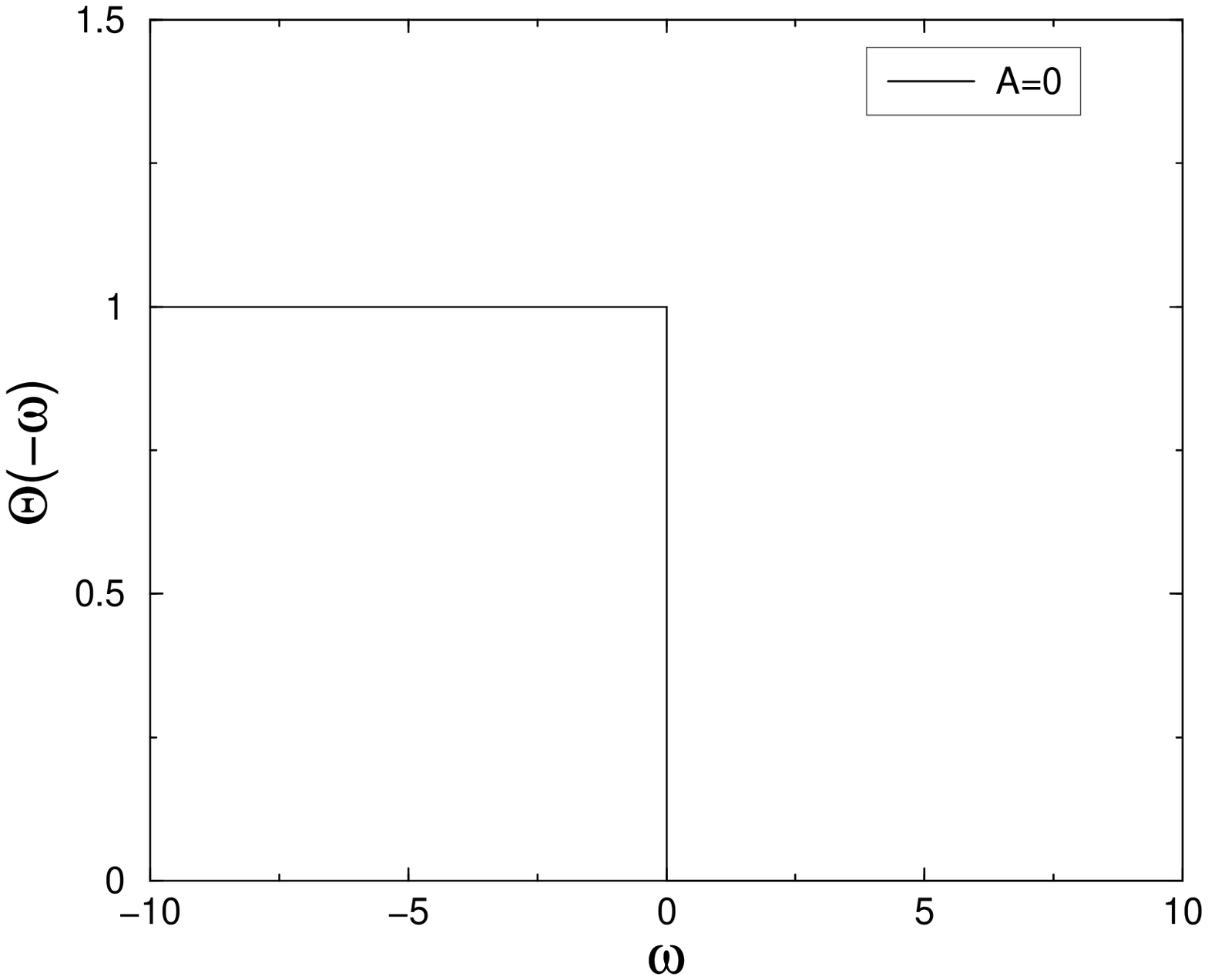, width=4cm, height=4cm}
\epsfig{file=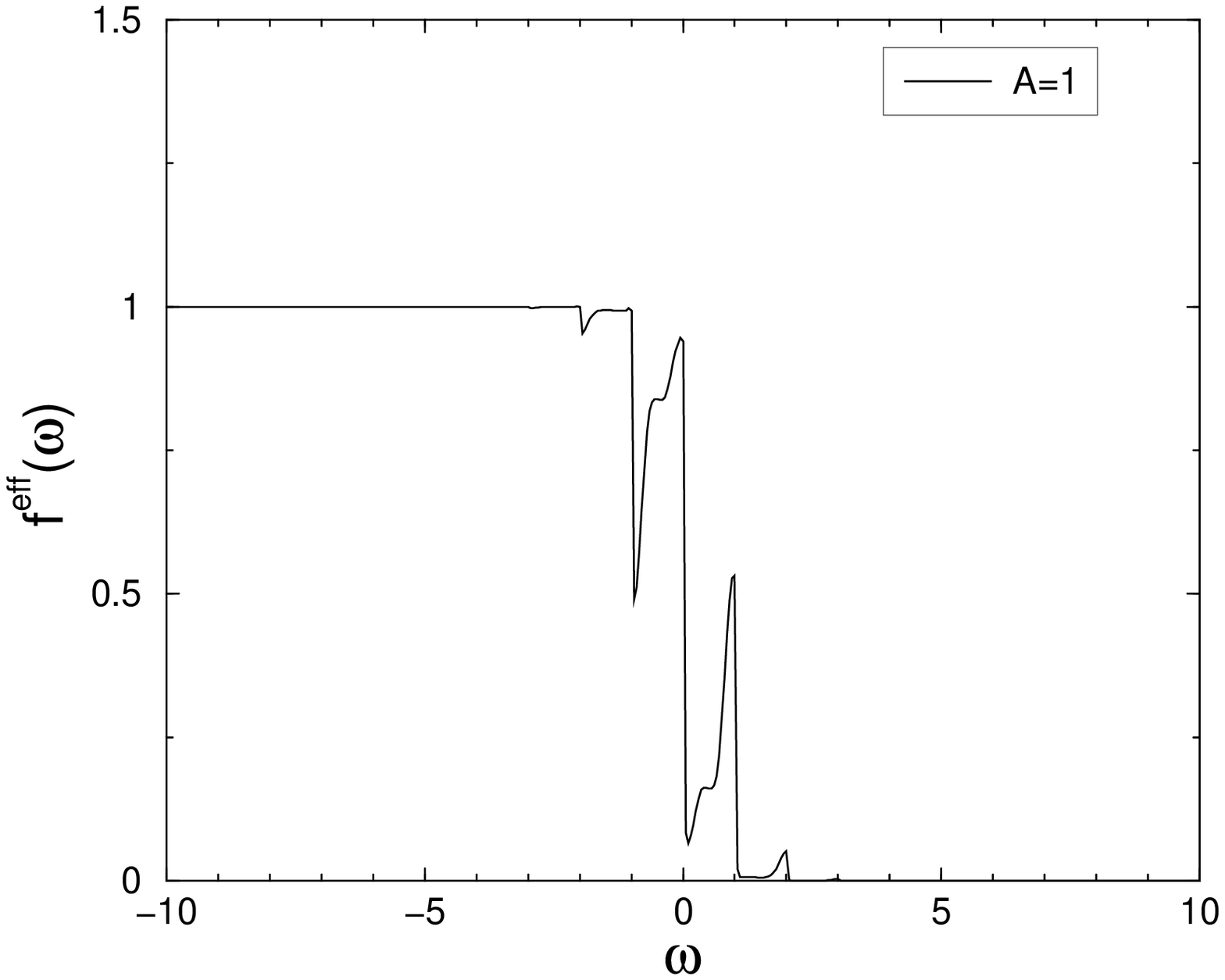, width=4cm,height=4cm}
\epsfig{file=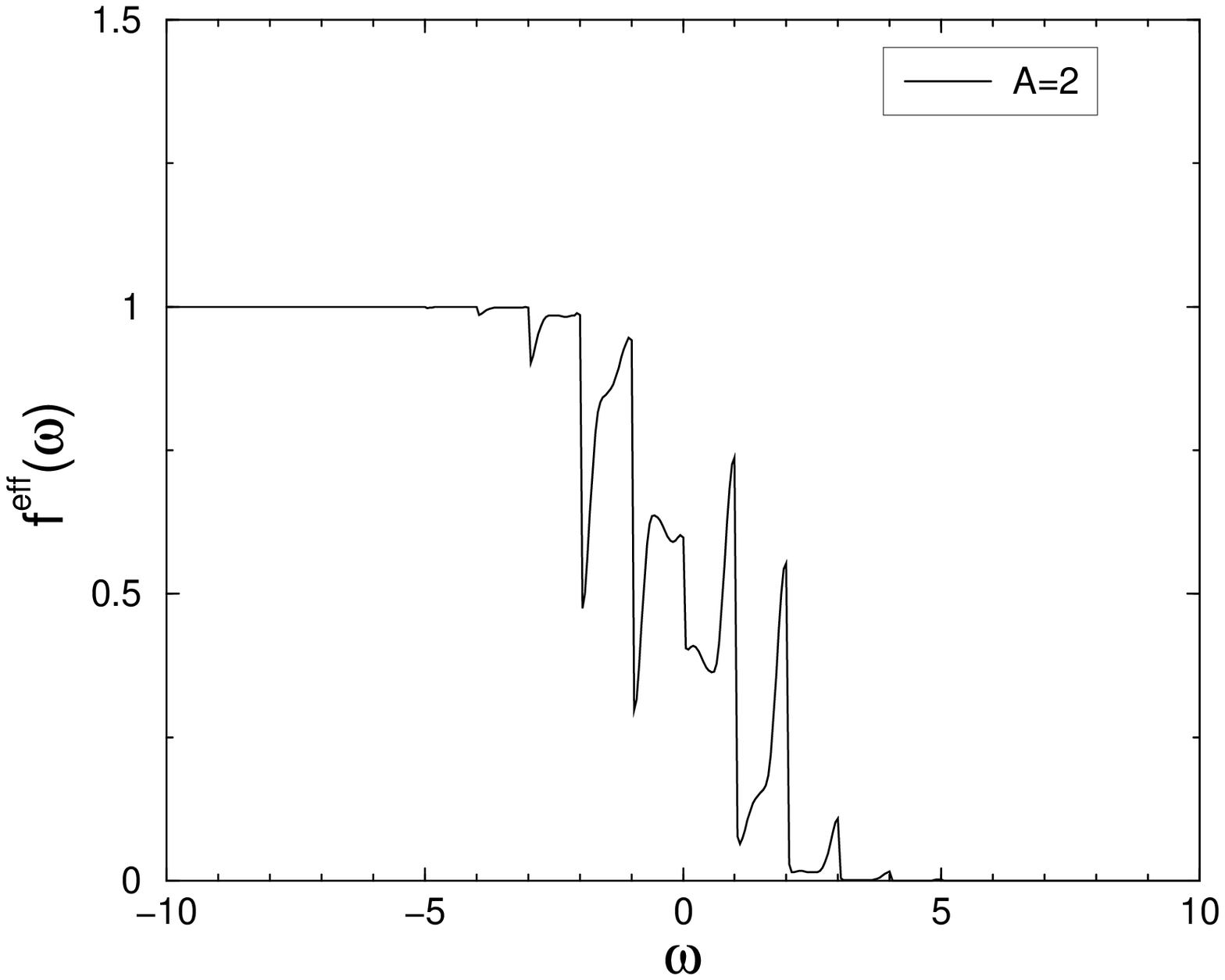, width=4cm,height=4cm}
\epsfig{file=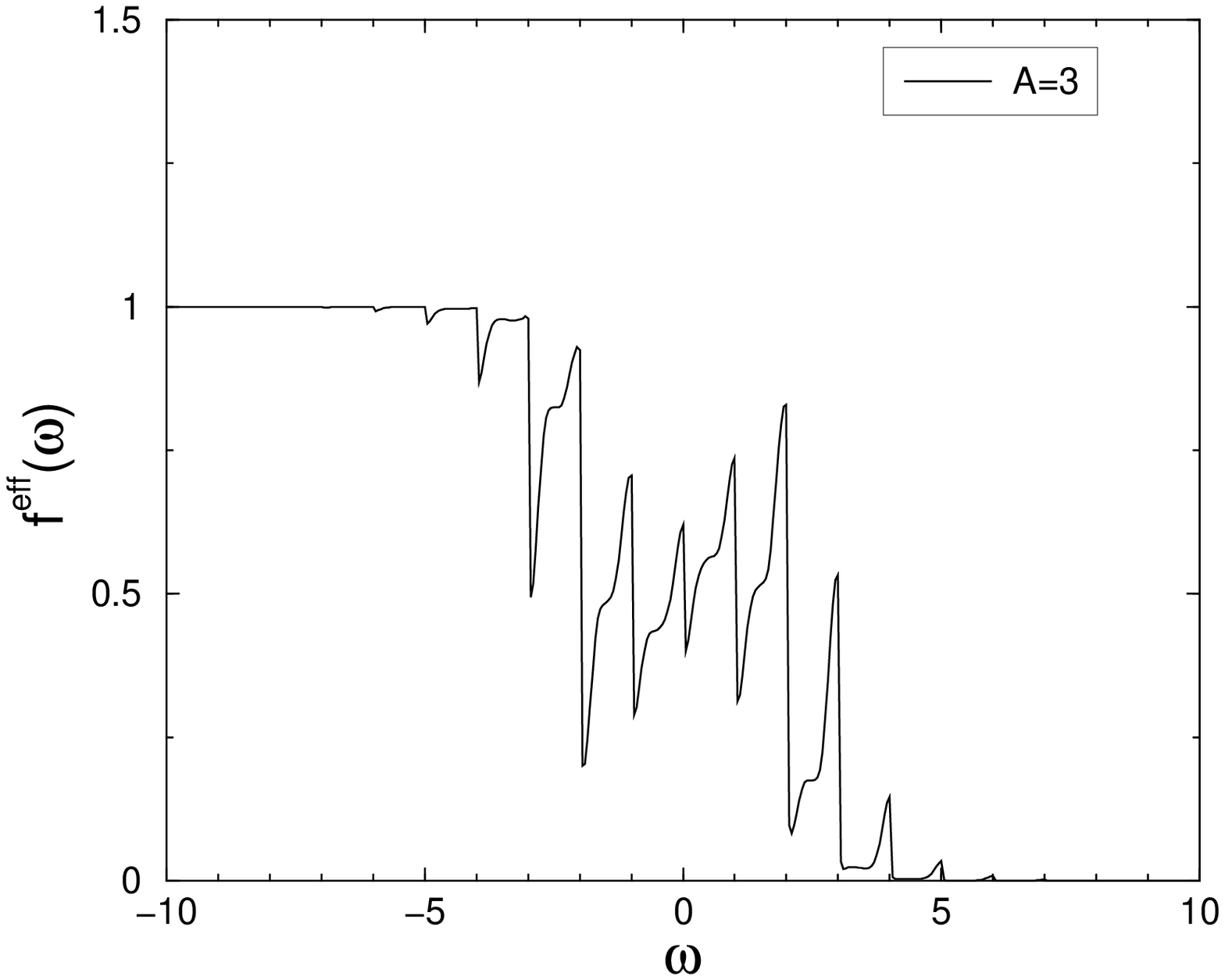, width=4cm, height=4cm}
\caption{Effective distribution function for $A=0,1,2,3$ at $T=0$.}
\label{FermiEff}
\end{center}
\end{figure}

Finally, we consider the Keldysh component of the Green's function, which contains information about the
occupation of the states. From $G_{00}^{\mbox{\tiny keld}}$ we define an effective distribution function:
\begin{equation}
G^{\mbox{\tiny keld}}_{00}(\omega) = -2\pi i(1-2f^{\mbox{\tiny eff}}(\omega))\rho^{\mbox{\tiny eff}}(\omega)
\end{equation} 
For zero temperature and $A=0$, $f^{\mbox{\tiny eff}}$ is the usual Heavi\-side function.
The effective distribution functions for $A=1,2,3$ are shown in Fig. \ref{FermiEff}.
The modifications due to the external field at the energy sidebands of the Fermi edge correspond to emission and absorption processes of energy quanta $\Omega$.
%----------------------------------------------------------------------
%
Experiments to measure the energy distribution function in a stationary out-of-equilibrium situation 
in metallic wires connected to two reservoir electrodes with an applied bias voltage have been done 
by Pothier et al. \cite{P}. 
%
%----------------------------------------------------------------------
%
%electrons at the Fermi edge can absorb the energy quantum $\Omega$ and electrons with energy $\epsilon_F-\Omega$ 
%can occupy the free states at the Fermi edge and absorb or emit further energy quanta. 
%At low field amplitude, $f^{\mbox{\tiny eff}}$ shows only one significant sideband of the Fermi edge but the number is enhanced
%with increasing field amplitude.
%
%
%Optical conductivity
%
%
The time-dependent DMFT also provides a useful method for calculating the optical conductivity of the driven Hubbard model.
%Linear-response theory for non\-equi\-li\-brium systems can be formulated quite simply in terms of Keldysh propagators.
Consider electrons described by the driven Hubbard model, which couple to an additional external classical vector potential
with wave vector ${\bf p}$. 
Linear-response theory for non\-equi\-li\-brium systems can be formulated quite simple in terms of Keldysh propagators. 
The paramagnetic current response in $\hat{\bf e}-$direction is given by
\begin{eqnarray}
 \left<J_{\hat{\bf e}}^{\mbox{\tiny param}}\left({\bf p},t\right)\right> & = & 
	\frac{et}{2}\sum_{{\bf k}\sigma}v_{\hat{e}}\left({\bf k}+{\bf p},{\bf k}\right)
	G_{{\bf k}+{\bf p}{\bf k}\sigma}^{\mbox{\tiny keld}}(t,t)%')\bigg|_{t=t'} \nonumber\\
%%%%%	& & +\ ie^2t\sum_{\mbf{k}\sigma}\cos k_\delta \ 
%%	G_{\mbf{k}\mbf{k}\sigma}^{\mbox{\tiny keld}}(t,t')A_\delta\left(\mbf{p},t\right)\bigg|_{t=t'},
\end{eqnarray}
with the vertex factor
\begin{equation}
 v_{\hat{\bf e}}\left({\bf k}+{\bf p},{\bf k}\right) = e^{i({\bf k}+{\bf p})\hat{\bf e}}-e^{-i{\bf k}\hat{\bf e}}.
\end{equation}
Just as in the case of the bare Hubbard Hamiltonian, all vertex corrections in the ladder decomposition of the current response
function vanish in the limit of infinite dimensions \cite{GKII,PCJI,PCJII,AK}. 
Thus, the optical conductivity is determined by the elementary particle-hole bubble as in the equilibrium case.
Assuming isotropic spatial conditions, we obtain for the paramagnetic contribution  
\begin{eqnarray}
 \sigma^{\mbox{\tiny param}}(\omega) & = & \frac{e^2}{2\omega d}\sum_{{\bf k}\sigma m}{\bf v}^2_{{\bf k}}\int\frac{d\omega'}{2\pi}\left[
	G^{(0),\mbox{\tiny keld}}_{{\bf k}\sigma 0m}(\omega'+\omega)%G^{(0),\mbox{\tiny av}}_{{\bf k}\sigma m0}(\omega')
	\right.\\
	& & \nonumber\\
	& \times & \left.G^{(0),\mbox{\tiny av}}_{{\bf k}\sigma m0}(\omega') + \ G^{(0),\mbox{\tiny ret}}_{{\bf k}\sigma 0m}(\omega'+\omega)\nonumber
	G^{(0),\mbox{\tiny keld}}_{{\bf k}\sigma m0}(\omega')\right],
\end{eqnarray}  
where $G^{(0)}$ denotes the zeroth order contribution in the classical vector potential, $\tau_1$ is the first Pauli matrix 
and ${\bf v_k}=\nabla_{\bf k}\epsilon_{\bf k}$.
%
%Assuming isotropic spatial conditions, 
%
Using the spectral representation of the Green's functions, the optical conductivity can be expressed in terms of 
the Fourier components of the spectral density $\rho^{l}_{0n}(\epsilon) = J_{l}(\frac{A}{\Omega})
J_{l-n}(\frac{A}{\Omega})\mbox{Im}\tilde{G}^{\mbox{\tiny av}}(\epsilon-l\Omega)/\pi$ and the free density of states $D(\epsilon)$
(the diamagnetic term cancels the $1/\omega$-divergence of the paramagnetic part in the optical conductivity). 
\begin{eqnarray}
 \sigma(\omega) & = & \frac{2 \pi e^2t^2}{\omega}\sum_{l_1,l_2,m}\int\limits_{-\infty}^{\infty}d\epsilon
	\int\limits_{-\infty}^{\infty}d\omega'
	D(\epsilon)\nonumber\\
	& & \times\quad\rho_{0m}^{l_1}(\epsilon,\omega'+\omega) \rho_{m0}^{l_2}(\epsilon,\omega') \nonumber\\
	& & \nonumber\\
	& & \times\quad\left(f\left(\omega'-l_2\Omega\right)-f\left(\omega'+\omega-l_1\Omega\right)\right)
\end{eqnarray}
This final result for $\sigma(\omega)$ significantly shows the appea\-rance of the energy sidebands due to the external driving 
field, and for zero driving field it reduces to the well-known expression for the optical conductivity of the bare Hubbard model.
%:
%
%\begin{eqnarray}   
% \mbox\sigma^{\mbox{\tiny Hubbard}}\left(\omega) & = & 
%	2 \pi e^2t^2\int\limits_{-\infty}^{\infty}d\epsilon\int\limits_{-\infty}^{\infty}d\nu D(\epsilon)\rho(\epsilon,\nu) \nonumber\\
%%% & & \times \rho(\epsilon,\nu+\omega)\frac{f(\nu)-f(\nu+\omega)}{\omega}.
%\end{eqnarray}
%
%Conclusions -----------------------------------------------------------------
%\section{ Conclusions }
%

We have generalized the DMFT equations using the Keldysh formalism to study a Hubbard model under the influence of a strong 
periodic external perturbation with arbitrary frequency and amplitude.
We found that the Green's functions were matrices in the energy and momentum sidebands due to quasi-energy and momentum 
conservation modulo the external frequency respectively wave vector.
 
We have derived a closed set of self-consistency equations for the driven Hubbard model wave vector nu\-me\-ri\-cally in
the special case of a uniform field using IPT. We obtained an effective spectral density and distribution function for
different values of the field amplitude. In \mbox{addition}, we have derived the frequency-dependent optical conductivity of the driven Hubbard model.

Solving the problem for a nonzero commensurate ${\bf q}$ would require an even larger numerical effort. One has to consider cluster
of $N = 2\pi/q$ sites and perform time-dependent DMFT for all sites ${\bf R}$ of this cluster.
This is basically the same as using dynamical cluster approximation (DCA) \cite{HTJ} in combination with the Keldysh for\-ma\-lism.
The original driven Hubbard model is mapped onto a self-consistent embedded cluster given by the periodic form of our Hamiltonian
instead of a single impurity. Decreasing the wave vector requires solving cluster problems of increasing size, which is numerically 
limited because of an exponentially growing Hilbert space. 

We are currently working on an application of the time-dependent DMFT to describe the local nonequilibrium physics of 
excitonic insulators and Mott insulators under laser excitation. This method shows great promise for being suitable to explain 
recent pump-probe experiments on transition metal oxides where unusual effects, e.g. a dramatic change in the optical 
transmission \cite{OA}, have been observed. The real-time development of the DOS and the gap after the impact of a short intense 
laser pulse is under investigation. 

%An interesting application of the time-dependent DMFT will be the investigation of optical properties, in particular coherent 
%many-body effects in the nonlinear optical response of modulation doped semiconductors \cite{PH,PSEC,Jauho}.
%Of particular interest are the spectral function, recombination rates and the photo conductivity as a function of the relevant 
%physical parameters, e.g. the temperature, the doping and the band width.
%
% References ------------------------------------------------------------------
%\bibliographystyle{unsrt}
%\bibliography{PetraDiplomArchive}
%

% End of document -------------------------------------------------------------
\end{multicols}
\end{document}